# Replication and study of anomalies in LK-99—the alleged ambient-pressure, room-temperature superconductor

T. Habamahoro[1,2], T. Bontke[1,2], M. Chirom[2,3], Z. Wu[1,2], J. M. Bao[2,3,4], L. Z. Deng[1,2*], C. W. Chu[1,2*]

[1]Department of Physics, University of Houston, Houston, Texas 77204, USA

[2]Texas Center for Superconductivity at the University of Houston (TcSUH), Houston, Texas 77204, USA

[3]Department of Chemistry, University of Houston, Houston, Texas 77204, USA

[4]Department of Electrical and Computer Engineering, Houston, Texas 77204, USA

*: ldeng2@central.uh.edu, cwchu@uh.edu

Abstract

We have studied LK-99 [$Pb_{10-x}Cu_x(PO_4)_6O$], alleged by Lee *et al.* to exhibit superconductivity above room temperature and at ambient pressure, and have reproduced all anomalies in electric and magnetic measurements that they reported as evidence for the claim of LK-99 being an ambient-pressure, room-temperature superconductor. We found that these anomalies are associated with the structural transition of the $Cu_2S$ impurity in their sample and not with superconductivity.

1. Introduction

Superconductivity is widely regarded as one of the greatest scientific discoveries of the 20[th] century. The unique simultaneous appearance of zero resistivity and magnetic field expulsion gives superconductors tremendous potential for various applications, such as efficient electric power transmission [1], much smaller or more powerful magnets, magnetic levitation, quantum computers, *etc.* Due to the century-old effort on superconductivity, the superconducting critical transition temperature ($T_c$) has continuously been enhanced. It appears that, since 1994, all record-high $T_c$s at various times have been achieved in compounds under very high pressure, *e.g.* 164 K in $HgBa_2Ca_2Cu_3O_{8+\delta}$ at 32 GPa in 1994 [2], 203 K in $H_3S$ at 150 GPa in 2015 [3], 260 K in $LaH_{10}$ at 190 GPa in 2019 [4], and 288 K in C-S-H at 267 GPa in 2020 (retracted in 2022) [5], as well as 294 K in N-doped $LuH_x$ at 1 GPa in 2023 [6]. One of the most serious challenges in meeting the goal of full deployment of superconductivity technology is to find a way to retain the coherent quantum state under ambient conditions, *i.e.* room temperature of ~ 300 K and atmospheric pressure.

On July 22, 2023, Lee *et al.* reported the observation of room-temperature superconductivity (RTS) with a $T_c$ ~ 400 K at ambient pressure in a material named LK-99 [7, 8], which is a Cu-doped lead phosphate apatite $Pb_{10-x}Cu_x(PO_4)_6O$. The main pieces of evidence for RTS in LK-99 provided were: 1) a video showing a piece of their sample partially levitating over a magnet at room temperature; 2) large diamagnetic susceptibility up to 350 K; 3) a diamagnetic transition near 400 K; 4) a sharp drop in resistance at ~ 400 K; and 5) sharp transitions in the current-voltage (I-V) characteristics at different temperatures up to ~ 400 K. Because pressure is not required in these experiments, many attempts to replicate the observations were carried out quickly in the weeks afterward. Almost all experiments reported to date, for example [9-12], refute the claim of superconductivity by Lee *et al.*, although numerous theoretical predictions with opposite conclusions have also appeared, such as [13-17].

Most of the experiments on LK-99 reported to date were done in haste during the first few weeks after the claim of ambient-pressure, room-temperature superconductivity was made on July 22, 2023. Given the significance of this claim, we decided to revisit the evidence reported. By studying samples prepared following the recipe of Lee *et al.* and a different route, we found: 1) the partial magnetic levitation between a magnet and their sample shown in the video mentioned above cannot be between a magnet and a superconductor; 2) the diamagnetic susceptibility is too unrealistically large to be true; 3) the diamagnetic transition near 400 K may be associated with a magnetic impurity phase of $Cu_2S$ in the LK-99 sample; 4) the sharp drop in resistance at ~ 400 K may be also be due to the resistive transition of the $Cu_2S$ impurity [9, 18]; and 5) sharp transitions in the I-V characteristics at different temperatures up to ~ 400 K are caused by a thermal switch in $Cu_2S$ and are not superconducting-normal transitions. Additionally, a sample not prepared according to Lee *et al.*, which does not contain $Cu_2S$, was found to exhibit very high resistivity and did not display any of the above anomalies. We therefore conclude with confidence that LK-99 is not a room-temperature superconductor at ambient pressure and the reported anomalies are attributed to phase transitions in the $Cu_2S$ impurity phase in the LK-99 sample.

## 2. Experimental section

2.1 Sample preparation

To examine the anomalies reported by Lee *et al.* as evidence for superconductivity in LK-99 [$Pb_{10-x}Cu_x(PO_4)_6O$], we prepared LK-99 samples following closely their reported recipe [8] *via* the solid-state reaction method, as well as a LK-99 sample without $Cu_2S$ by following a slightly different path for comparison. The Lanarkite [$Pb_2(SO_4)O$] precursor was produced by reacting lead oxide powder (PbO, Alfa Aesar, 99.99%) with lead sulfate powder ($PbSO_4$, prepared from

PbO and sulfuric acid, $H_2SO_4$) in a molar ratio of 1:1 at 725 °C in air or in vacuum as mentioned by Lee *et al.* [8]. It should be noted that the Lanarkite was white when formed in vacuum while light yellow when formed in air. The lattice parameters for the latter appear to contract slightly in comparison with the former, as evidenced from their respective x-ray diffraction (XRD) patterns (Fig. 1). The copper phosphide ($Cu_3P$) precursor was produced by sintering a stochiometric mixture of copper powder (Alfa Aesar, 99.99%) and phosphorus lumps (Aldrich, 99.999%) at 540 °C for 24 hours in an evacuated quartz tube. These two precursors were thoroughly mixed with P in different molar ratios, and each mixture was pelletized, loaded in an alumina crucible, sealed in an evacuated quartz tube at $10^{-5}$ torr, and heated at 925 °C for 10 hours. Three samples were prepared with molar ratios of $Pb_2(SO_4)O:Cu_3P:P$ = 3:3:1 for S1 [$Pb_{10-x}Cu_x(PO_4)_6O$]; 1:1:0 for S2 [$Pb_{10-x}Cu_x(PO_4)_6O$]; and 5:0:6 first to make lead phosphate apatite [$Pb_{10}(PO_4)_6O$], which was then mixed with Cu in a 1:1 molar ratio to form S3 [$Pb_{10-x}Cu_x(PO_4)_6O$]. We also sintered $Cu_2S$ for later analysis to confirm that it is the impurity phase in the LK-99 sample reported by Lee *et al.* The $Cu_2S$ powder (Alfa Aesar, 99.99%) was pressed into a pellet, sealed in an evacuated quartz tube at $10^{-5}$ torr, and heated at 925 °C for 10 hours. The XRD patterns for S1-3, LK-99 reported by Lee *et al.* [8], and $Cu_2S$ are shown in Fig. 2. Qualitatively, the main phases in samples S1-3 are the same as those in LK-99. In addition, S1, S2, and LK-99 all display the $Cu_2S$ impurity, but S3 does not, as highlighted in the region between the dashed lines in Fig. 2.

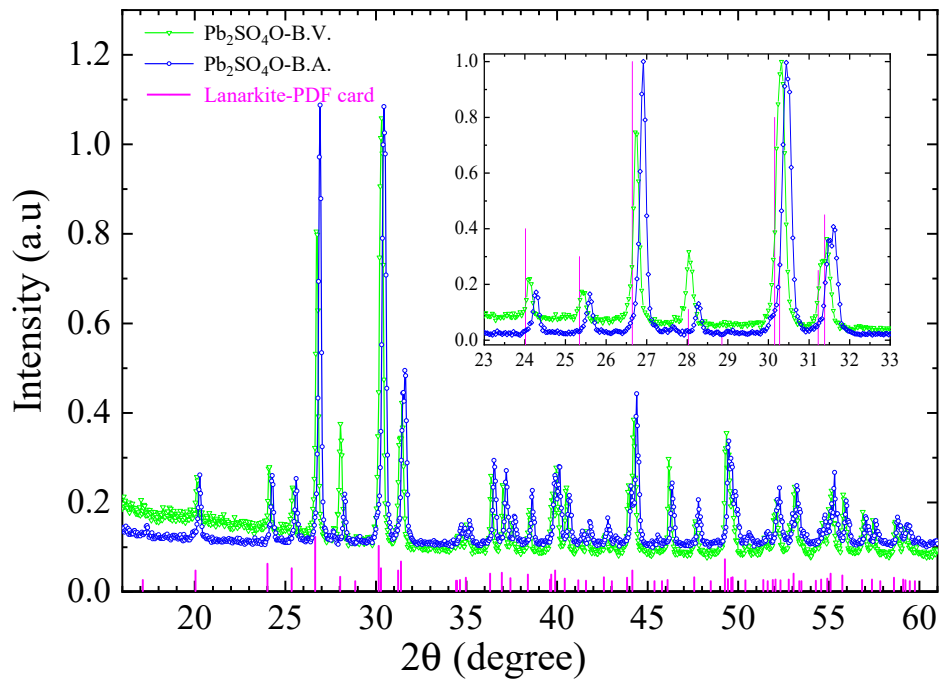

Fig. 1. XRD patterns of the Lanarkite $Pb_2(SO_4)O$ precursors in comparison with reference (Lanarkite PDF card #33-1486). $Pb_2(SO_4)O$-B.V. and $Pb_2(SO_4)O$-B.A. represent Lanarkite samples synthesized in vacuum and in air, respectively. Inset: expanded view clearly showing lattice shrinkage when Lanarkite is synthesized in air.

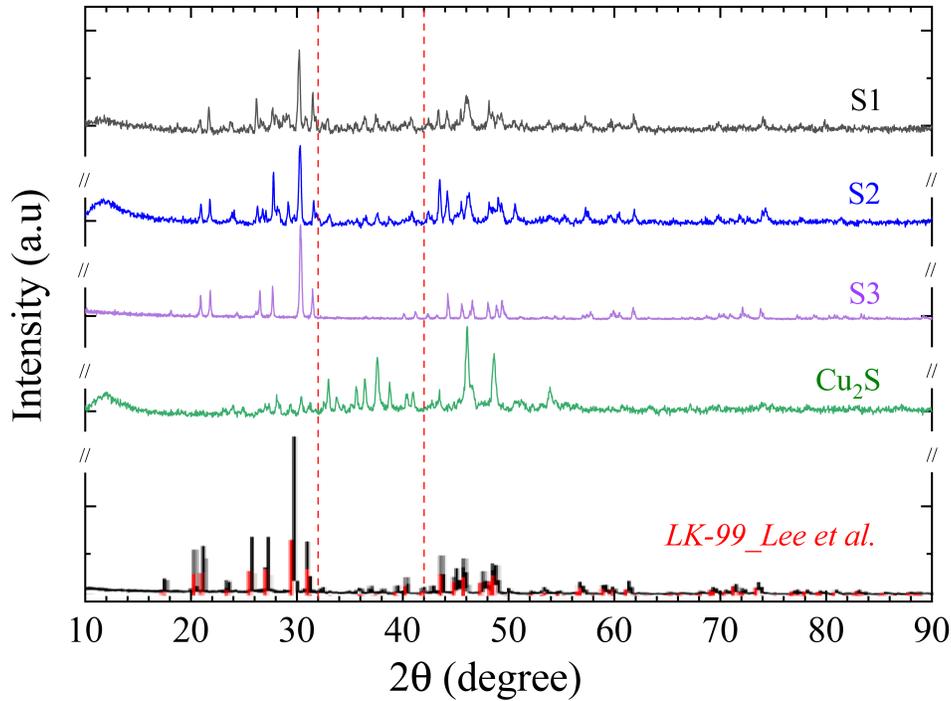

Fig. 2. XRD patterns of samples S1 (black), S2 (blue), and S3 (violet); pure $Cu_2S$ (green); and LK-99 (black) and lead apatite (red) reported by Lee *et al.* (reproduced from [8]).

2.2 X-ray measurements

The crystal structure and phases of the produced samples were verified using an x-ray powder diffractometer (Rigaku) equipped with a CuKa1 x-ray source ($\lambda = 1.5406$ Å).

2.3 Electrical transport measurements

The electrical resistance of each sample as a function of temperature was measured by employing a standard 4-probe method in a Physical Property Measurement System (PPMS, Quantum Design). The voltage as a function of current for S1 and pure $Cu_2S$ were measured using a homemade probe with a Keithley 2400 current source and a Keithley 2182 nanovoltmeter. A temperature sensor was mounted on the sample to monitor the temperature changes. The resistivity of the highly insulating sample S3 was determined at room temperature using a Keithley 6517A electrometer/high resistance meter.

2.4 Magnetization measurements

DC magnetic susceptibility measurements were performed using a Magnetic Property Measurement System (MPMS, Quantum Design).

## 3. Results and Discussion

Except for the partial magnetic levitation, all anomalies at 350-400 K reported by Lee *et al.* in LK-99 have been qualitatively reproduced in our S1, S2, and $Cu_2S$ samples, but not in S3. They are shown and discussed individually below.

3.1 The magnetic levitation

Under proper conditions, the observation of directional-free magnetic levitation of a sample material above a magnet provides the best evidence for the material's superconductivity. Unfortunately, the video provided by Lee *et al.* shows that their wedge-shaped sample was only partially levitated, with the heavy end anchored on the magnet and with the plane following the magnetic flux line flaring out from the center of the magnet, as was evident when the sample was pushed. Stable magnetic levitation of a strong diamagnet above a magnet is not expected based on Earnshaw's theorem of interacting magnetic objects. As demonstrated in Fig. 3, a thin diamagnetic graphite disk can only be levitated stably by an assembly of magnets with a special field profile. It should also be noted that a Type-I superconductor cannot be levitated stably by a magnet without the intervention of an external magnetic field, but a Type-II superconductor can be stably levitated by a magnet against its field direction due to the trapped magnetic flux, as commonly seen for the YBCO high-temperature superconductor.

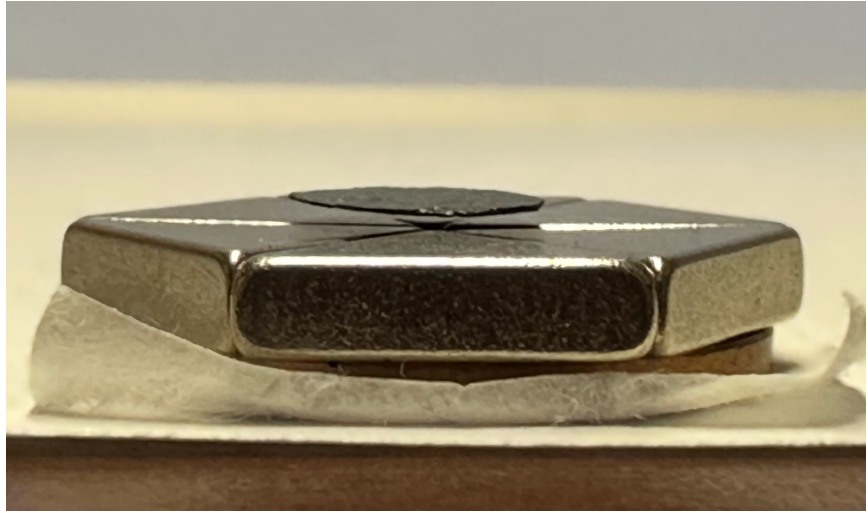

Fig. 3. Stable magnetic levitation of a thin slab of graphite over an assembly of magnets with a specific field profile (by T. C. Chiang of UIUC).

3.2 The large diamagnetic susceptibility

A perfect diamagnetic susceptibility of $-1/4\pi$ is the hallmark of a superconductor below its $T_c$. However, the value reported by Lee *et al.* was unphysically large, even larger than $-1/4\pi$. The magnetic susceptibility of our sample S1 shown in Fig. 4a is much smaller than that published (Fig. 4 of [8]).

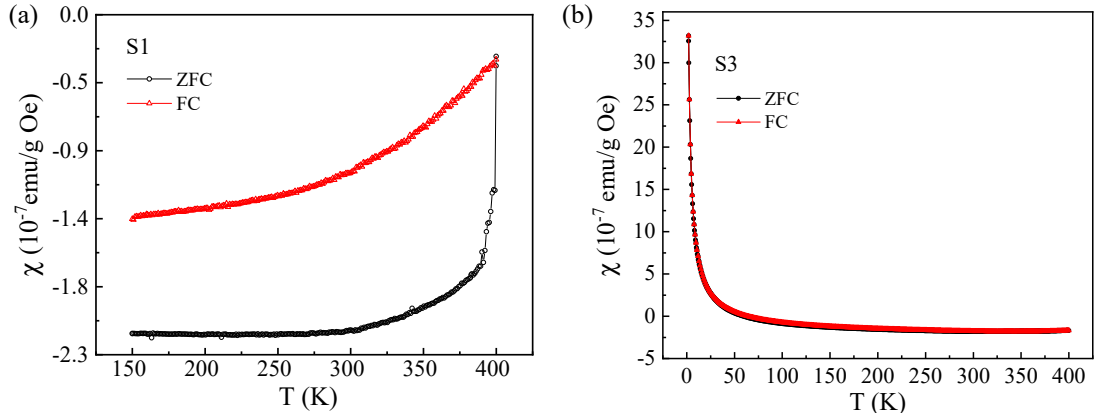

Fig. 4. Magnetic susceptibility as a function of temperature for (a) S1, showing a sharp diamagnetic transition at around 380 K, and (b) S3, showing no sign of transition around 380-400 K.

3.3 The diamagnetic shift near 400 K

The reported diamagnetic shift near 400 K under ZFC measurement observed in LK-99 was detected in our sample S1 that contains the $Cu_2S$ impurity, as shown in Fig. 4a, but not in our sample S3, in which $Cu_2S$ was absent, as shown in Fig. 4b.

3.4 The resistance drop near 400 K

The resistance drop near 400 K reported by Lee *et al.* for LK-99 was observed in our sample S1 with the $Cu_2S$ impurity, but not to zero, as shown in Fig. 5a. It is interesting to note that, although a resistance anomaly was also detected over a similar temperature range for sample S2 (Fig. 5b), the S2 resistance decreased with increasing temperature, opposite to the trend for S1. This might result from differences in their synthesis conditions, such as S1 having more P. Further studies are beyond the scope of this paper.

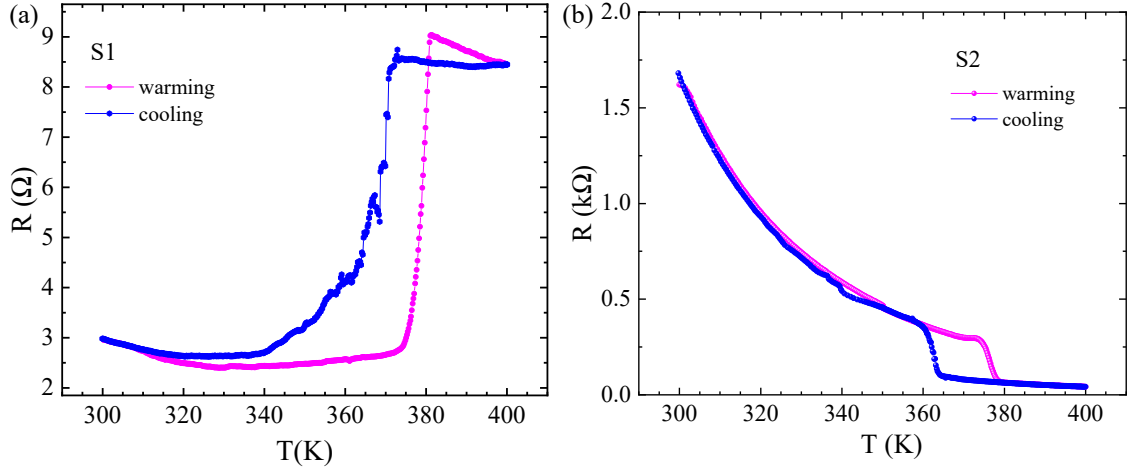

Fig. 5. Resistance *vs.* temperature from 300 K to 400 K for (a) S1 and (b) S2.

3.5 The sharp current-induced switches in the I-V characteristics at different temperatures

The current-induced switches in the I-V characteristics of LK-99 reported by Lee *et al.* were reproduced in our sample S1, as shown in Fig. 6, but not in sample S3. We also measured simultaneously the temperature of sample S1 as a function of I and found that both its temperature and its resistance increase with increasing I. When the temperature reaches ~ 328 K (it should be noted that the temperature sensed by the thermometer was lower than that of the sample due to the thermal lag between the sample and the thermometer), the resistance rises rapidly, signaling a transition from the low-temperature monoclinic to high-temperature hexagonal structure of $Cu_2S$. Based on this observation, we conclude that the current-induced anomaly in I-V characteristics represents a current-induced thermal switch rather than the current-induced superconducting-normal transition proposed by Lee *et al.*

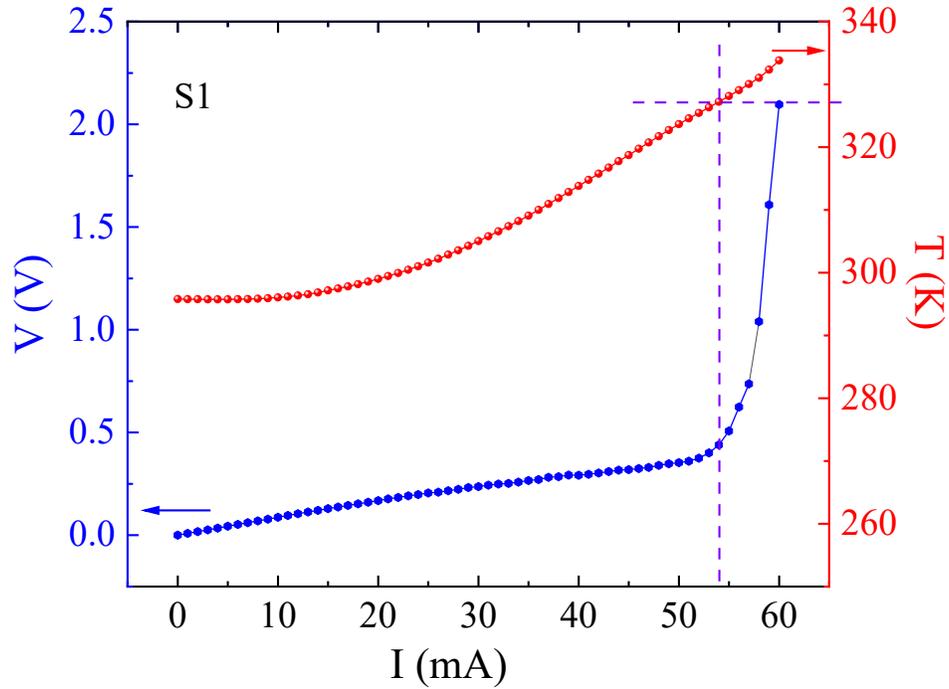

Fig. 6. Voltage (blue) and temperature (red) *vs.* current for S1. Temperature was monitored by a sensor in contact with the sample.

3.6 Resistance of Sample S3 without $Cu_2S$

Demonstrating that the reported anomalies are associated with the impurity phase of $Cu_2S$, we found that the resistivity of our sample S3 without $Cu_2S$ is huge, $> 10^{11}$ Ω·cm, and that it does not exhibit any of the above-mentioned anomalies.

3.7 Properties of Pure $Cu_2S$

To show that the anomalies discussed above arise from $Cu_2S$, we measured our sintered sample, and detected in it the reported magnetic (Fig. 7a) and resistive (Fig. 7b) transitions in the temperature range of ~ 375-400 K, as well as the transition in its I-V characteristics, as shown in Fig. 7c.

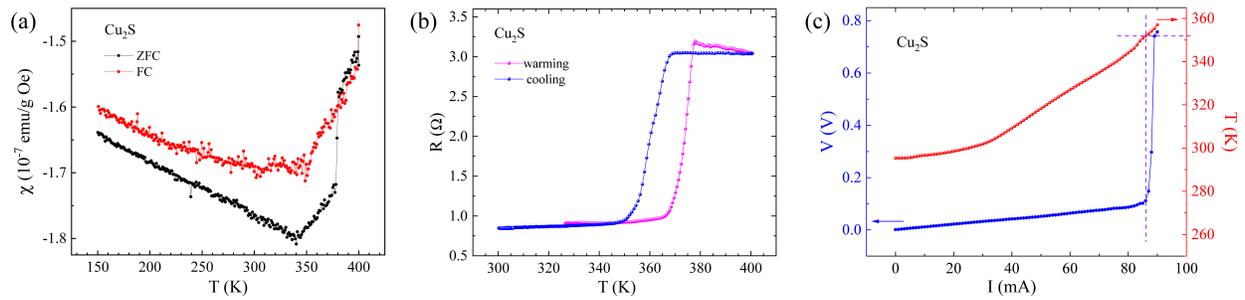

Fig. 7. Temperature-dependent (a) magnetic susceptibility and (b) resistance of $Cu_2S$, both showing a hysteresis at ~ 375 - 400 K. (c) Voltage (blue) and temperature (red) *vs.* current, showing a sudden jump due to a structural phase transition.

## 4. Conclusion

We have synthesized LK-99 [$Pb_{10-x}Cu_x(PO_4)_6O$] samples following closely the published recipe by Lee *et al.* and have reproduced qualitatively the reported anomalies in electric and magnetic measurements above room temperature at ambient pressure. Careful systematic characterization of these samples, of a pure LK-99 sample without the $Cu_2S$ impurity, and of pure $Cu_2S$ led us to the conclusion that anomalies considered by Lee *et al.* to be evidence for room-temperature superconductivity in LK-99 at ambient pressure are associated with the structural transition in the $Cu_2S$ impurity phase in their sample and not with a superconducting transition.

## References

[1]    Chu C W and Jacobson A J 2004 HTS/LH2 Super-Grid: An energy superhighway in the U.S. for the new millennium *J. Phys. Conf. Ser.* **181**, 67

Acknowledgements:

T.H., T.B., Z.W., L.Z.D., and C.W.C. are supported by the Enterprise Science Fund of Intellectual Ventures Management, LLC; U.S. Air Force Office of Scientific Research Grants FA9550-15-1-0236 and FA9550-20-1-0068; the T. L. L. Temple Foundation; the John J. and Rebecca Moores